\documentclass[prd,
,superscriptaddress,
nofootinbib,%
tightenlines ]{revtex4}
\usepackage{epsfig}
\usepackage[colorlinks=true,linkcolor=blue,urlcolor=blue,citecolor=blue]{hyperref}
\usepackage{amsmath}
\usepackage{amsfonts}
\usepackage{amssymb}
\usepackage{bm}

\newcommand{\ben}{\begin{displaymath}}
\newcommand{\een}{\end{displaymath}}
\newcommand{\be}{\begin{equation}}
\newcommand{\ee}{\end{equation}}
\newcommand{\bea}{\begin{eqnarray}}
\newcommand{\eea}{\end{eqnarray}}

\newcommand{\nn}{\nonumber \\ }

\begin{document}
\title{Definition of gravitational local spatial densities for spin-0 and spin-1/2 systems}
 \author{J.~Yu.~Panteleeva}
  \affiliation{Institut f\"ur Theoretische Physik II, Ruhr-Universit\"at Bochum,  D-44780 Bochum,
 Germany}
\author{E.~Epelbaum}
 \affiliation{Institut f\"ur Theoretische Physik II, Ruhr-Universit\"at Bochum,  D-44780 Bochum,
 Germany}
\author{J.~Gegelia}
 \affiliation{Institut f\"ur Theoretische Physik II, Ruhr-Universit\"at Bochum,  D-44780 Bochum,
 Germany}
 \affiliation{High Energy Physics Institute, Tbilisi State
University, 0186 Tbilisi, Georgia}
\author{U.-G.~Mei\ss ner}
 \affiliation{Helmholtz Institut f\"ur Strahlen- und Kernphysik and Bethe
   Center for Theoretical Physics, Universit\"at Bonn, D-53115 Bonn, Germany}
 \affiliation{Institute for Advanced Simulation, Institut f\"ur Kernphysik
   and J\"ulich Center for Hadron Physics, Forschungszentrum J\"ulich, D-52425 J\"ulich,
Germany}
\affiliation{Tbilisi State  University,  0186 Tbilisi,
 Georgia}

\date{10 July 2023}
\begin{abstract}
We work out details of defining the spatial densities corresponding to the
gravitational form factors of spin-0 and spin-1/2 systems using spherically symmetric 
sharply localized wave packets.  The expressions for the spatial densities are provided in the
frames with both zero and non-zero expectation values of the momentum operator.

\end{abstract}

\maketitle

\section{Introduction}

In  close analogy with the electric charge density
of hadrons \cite{Hofstadter:1958,Ernst:1960zza,Sachs:1962zzc}, the interpretation of Fourier transforms of
gravitational form factors in the Breit frame as local densities of various physical quantities characterizing the system was suggested in Refs.~\cite{Polyakov:2002wz,Polyakov:2002yz,Polyakov:2018zvc}.
For systems whose
intrinsic size is comparable with their Compton wavelength,
this identification of spatial density distributions with the Fourier
transforms of the electromagnetic and gravitational form factors was criticized in
Refs.~\cite{Burkardt:2000za,Miller:2007uy,Miller:2009qu,Miller:2010nz,Jaffe:2020ebz,Miller:2018ybm,Freese:2021czn}.   
This issue has raised much interest recently
\cite{Lorce:2020onh,Lorce:2022cle,Guo:2021aik,Lorce:2018egm,Freese:2021mzg,Chen:2022smg,Freese:2022fat,Carlson:2022eps}.
The light-front approach has been used to
    define purely intrinsic electromagnetic spatial densities as two-dimensional
 distributions in the impact parameter space 
\cite{Burkardt:2000za, Miller:2007uy, Miller:2009qu, Miller:2010nz, Guo:2021aik}. 
 The relationship between
 these densities and the non-relativistic three-dimensional distributions 
 in the Breit frame in terms of the Abel transform was studied in
 Refs.~\cite{Panteleeva:2021iip,Freese:2021mzg}.
 The phase-space approach of Refs.~\cite{Lorce:2020onh,Lorce:2022cle,Lorce:2018egm} 
allows one to define fully relativistic three-dimensional spatial
densities. However, these densities do not have a strict probabilistic
 interpretation due to their dependence on both  coordinates 
 and momenta. 
In Refs.~\cite{Freese:2021czn, Lorce:2018egm}, the two-dimensional light-front Fourier transforms of the  energy-momentum tensor 
form factors have been interpreted 
as the spatial densities of the energy-momentum and the internal forces in hadrons.   
In Ref.~\cite{Epelbaum:2022fjc}, a definition of spatial densities of local operators for systems with arbitrary Compton wavelengths 
has been suggested  by considering an example of the charge density   
of a spin-$0$ system,  see also Ref.~\cite{Fleming:1974af} for an early study along this line.  
Recently this novel definition was also applied to the electromagnetic spatial densities for spin-1/2 systems \cite{Panteleeva:2022khw}. 
For illuminating comments and extensions 
on the approach using sharply localized packets
see Ref.~\cite{Carlson:2022eps}. 

In this work we consider the one-particle matrix elements of the energy-momentum tensor (EMT) in localized states of spin-0 and spin-1/2 systems. 
 Specifying the one-particle state by a spherically symmetric wave packet and sharply localizing it in space 
 we define  spatial distributions corresponding to gravitational form factors in the zero average momentum frame (ZAMF). Next, we generalize the new definition to Lorentz-boosted frames. 
Analogously to Ref.~\cite{Jaffe:2020ebz}, which dealt with the electromagnetic case, we also consider the static approximation and discuss the limitations of the results obtained in this limit. 

Recently in Ref.~\cite{Freese:2022fat} it has been claimed that the idea of obtaining internal densities via wave packet localization is fallacious, because it does not reproduce the ``true internal densities''. 
The authors first clarify what they understand under true internal densities of hadrons and how these are related to the physical densities obtained from expectation values of local operators. 
Next, using the formulated criteria they observe that the wave packet localization does not result in meaningful internal densities for instant form coordinates.   
We believe that the parameterization of the physical densities in terms of internal densities and smearing functions which is suggested in Ref.~\cite{Freese:2022fat} as the basis 
for defining the true internal densities has an oversimplified form and is inadequate for a complicated hadronic system in a state described by a wave packet.

Our work is organized as follows.
In Sec.~\ref{GrD} we define spatial densities of the EMT in the ZAMFs of a spin-0 and a spin-1/2 systems and consider the static approximation. Spatial densities in moving reference frames are 
considered in Sec.~\ref{MFrames}. Section~\ref{interpretation} contains our interpretation of the obtained results and we summarize in Sec.~\ref{summary}.

\section{Gravitational densities in the zero average momentum frame}
\label{GrD}

Spatial densities of the EMT defined via localized states differ significantly from the ones of the electromagnetic current.  A superposition of eigenstates of the charge operator, 
which makes the localized packet, remains an eigenstate of the charge operator with the same eigenvalue. However this is not the case for the energy-momentum operator.  That is, a packet which is a superposition 
of one-particle eigenstates of the four-momentum operator with different four-momenta is not an eigenstate of the same operator. 
On the other hand, such a state does not contain admixtures of states with 
particle-antiparticle pairs and, therefore, can be used to define spatial distributions for one-particle states. 

Below we work out details of the spatial densities corresponding to the EMT of  spin-0 and spin-1/2 systems in the ZAMF and in moving frames.
Throughout this work we closely follow the notations of Ref.~\cite{Jaffe:2020ebz}. 
 We choose the four-momentum eigenstates
$|p,s\rangle$ characterizing our system to be normalized as
\begin{equation}
\langle p',s'|p,s\rangle = 2 E (2\pi )^3 \delta_{s's}\delta^{(3)} ({\bf p'}-{\bf p})\,,
\label{NormStateN}
\end{equation}
where $(p,s)$ and $(p',s')$ are momentum and polarization of the initial and final states of a spin-1/2 system, respectively.  
For spin-0 systems we have analogous expressions without polarizations $s$ and $s'$ (this is also the case   
for the expressions below, which are applicable to both spin-0 and spin-1/2 systems). Further, we have  $p=(E,{\bf p})$, with $E=\sqrt{m^2+{\bf p}^2}$, where $m$ is the particle mass.

When calculating matrix elements of the EMT we use normalizable Heisenberg-picture states written in terms of wave packets as follows:
\begin{equation}
|\Phi, {\bf X},s \rangle = \int \frac{d^3 {p}}{\sqrt{2 E (2\pi)^3}}  \, \phi(s,{\bf p}) \, e^{-i {\bf p}\cdot{\bf X}} |p ,s \rangle,  
\label{statedefN2}
\end{equation}
where ${\bf X}$ is the spatial translation vector whose interpretation will be discussed below and $\phi(s,{\bf p})$ is the profile function satisfying the normalization condition
\begin{equation}
\int d^3 {p} \,  | \phi(s,{\bf p})|^2 =1\,.  
\label{normN}
\end{equation}
To {\it define} the density distributions in
the ZAMF of the system we use spherically symmetric wave packets.\footnote{The average momentum of the system vanishes in states corresponding to such packets.}  
In the case of spin-1/2 systems, the profile functions are also chosen to be spin-independent, i.e.~$\phi(s,{\bf p})=\phi({\bf p})=\phi(|{\bf p}|)$. 
It is convenient to define dimensionless profile functions 
\begin{equation}
\phi({\bf p}) = R^{3/2} \, \tilde \phi(R  {\bf p})\,,
\label{packageFormN}
\end{equation} 
where $R$ specifies the size of the wave packet. A sharp localization of the system
is achieved by taking small values of $R$.

\subsection{Spin-0 particles}

To define spatial densities associated with the EMT of a scalar particle we consider its matrix element in a state specified by Eq.~(\ref{statedefN2}) for a spin-0 state 
and take its limit when $R\to 0$.
Using the parametrization of matrix elements of the EMT in single-particle momentum eigenstates in terms of the form factors $\Theta_1(q^2)$ and $\Theta_2(q^2)$  \cite{Donoghue:1991qv,Kubis:1999db} 
we obtain: 
\begin{align}
t^{\mu\nu}_{\phi}( {\bf x} - {\bf X}) & = \langle \Phi, {\bf X} |\hat{T}^{\mu\nu}({\bf x},0)| \Phi, {\bf X} \rangle  = \int \frac{d^3 {p'} d^3 {p}}{(2\pi)^3 \sqrt{4 E' E }}  \, \phi^\star({\bf p'}) \,  \phi({\bf p})  
\langle p' |\hat{T}^{\mu\nu}({\bf x},0) |p \rangle \nn
& =
 \int \frac{d^3 {P} \, d^3 q}{(2\pi)^3 \sqrt{4 E
    E'}}\, \bigl[  \left( q^2 g^{\mu\nu}-q^\mu q^\nu \right) \Theta_1\left(q^2\right)+2 P^\mu P^\nu \Theta_2\left(q^2\right) 
      \bigr] 
                \phi\bigg( {\bf P} -
\frac{\bf q}{2} \bigg) \, \phi^\star\bigg( {\bf P} +\frac{\bf q}{2} \bigg)  \, e^{ -i {\bf q}\cdot ({\bf  x} - {\bf  X})}\,,
\label{rhoint2G}
\end{align}  
where we have introduced new variables $q = p'-p$ and $P=(p+p')/2$. In terms of these variables the energies are given as $E=(m^2+ {\bf P}^2 - {\bf P}\cdot {\bf q} +{\bf q}^2/4)^{1/2} $
and $E'=(m^2+ {\bf P}^2 + {\bf P}\cdot {\bf q} +{\bf q}^2/4)^{1/2} $.

By applying the method of dimensional counting of Ref.~\cite{Gegelia:1994zz}, the leading contribution in Eq.~(\ref{rhoint2G}) for $R\to 0$ can be obtained without 
performing the integration over momenta and without specifying the expressions for the form-factors and the spherically symmetric profile function. 
 For the form factors $\Theta_1\left(q^2\right)$ and $\Theta_2\left(q^2\right)$
  decaying for large $q^2$ as $1/q^4$ and $1/q^2$, or faster, respectively,\footnote{
 In Ref.~\cite{Tanaka:2018wea} using perturbative QCD
 the large-$q^2$ behavior of the gravitational form factors of the (pseudo)scalar hadrons has been
 found to be $\Theta_1\left(q^2\right)\sim 1/q^4$ and $\Theta_2\left(q^2\right)\sim 1/q^2$. 
 However, according to Ref.~\cite{Tong:2021ctu} in case of pions perturbative QCD leads to 
 $\Theta_1\left(q^2\right)\sim 1/q^2$, $\Theta_2\left(q^2\right)\sim 1/q^2$ for large $q^2$. 
 Notice that our derivation would not be applicable for the latter case.}
  the only non-vanishing contribution for $R\to 0$  
 originates from the large ${\bf P}$ region. This contribution can be obtained by substituting ${\bf P}= \tilde {\bf P}/R$ and expanding the resulting integrand in Eq.~(\ref{rhoint2G}) in powers of $R$
around $R=0$. Doing so and introducing a new variable ${\bf  r} = {\bf  x}-{\bf  X}$ we obtain 
\begin{eqnarray}
t^{\mu\nu}_\phi ( {\bf r}) & = & \int \frac{d^3 \tilde {P}  d^3 {q}}{(2\pi)^3}  \left( 
\frac{\tilde P^\mu \tilde P^\nu}{\tilde P R} \,  \Theta_2 \left[  \tilde q^2
\right]  
+\frac{R}{2 \tilde P} \left( 
\tilde q^2 g^{\mu\nu} - \tilde q^\mu \tilde q^\nu  \right)
\Theta_1 \left[ \tilde q^2
 \right] 
\right) 
 |\tilde\phi(\tilde {\bf P})|^2  \, e^{ - i {\bf q}\cdot {\bf  r}}+\text{Rest},
\label{rhoint3G}
\end{eqnarray}
where  we introduced four-vectors $\tilde q^\mu =( {\bf \hat{\tilde P}} \cdot{\bf q},{\bf q})$  and 
$\tilde P^\mu=(\tilde P, {\bf \tilde P})$ with $\tilde P = |{\bf \tilde P}|$ and $ {\bf \hat{\tilde P}}   = {\bf{\tilde  P}}/\tilde P$. 
Notice that we wrote explicitly only terms that contribute to the final expressions for the
  densities, see the discussion in Sec.~\ref{interpretation}, and "Rest" stands for all other terms of the expansion.
Using Eq.~(\ref{normN})  and spherical symmetry we arrive at the final form
of the local density distributions in the ZAMF of the system
\begin{align}
t^{\mu\nu}( {\bf r}) & = N_{\phi,R} \, 
\int \frac{d^2 \hat {\tilde P}  \, d^3 {q}}{(2\pi)^3} \, \hat {\tilde P}^\mu \hat {\tilde P}^\nu\, \Theta_2\left[  - {\bf q}_\perp^2  \right] e^{ - i {\bf q}\cdot {\bf  r}} 
- N_{\phi,R,2} \, \int \frac{d^2 \hat {\tilde P}  \,d^3 {q}}{(2\pi)^3} \,  
 \left( \tilde q^\mu \tilde q^\nu + {\bf q}_\perp^2  g^{\mu\nu} \right)\, 
\Theta_1 \left[ -{\bf q}_\perp^2 \right]    \, e^{ - i {\bf q}\cdot {\bf  r}}+\text{Rest}
\,,
\label{rhoint4G}
\end{align}
where  
$\hat {\tilde P}^\mu =\left(1,{\bf \hat{\tilde P}}\right)$, ${\bf q}_\perp = {\bf \hat{\tilde P}} \times \left({\bf q} \times {\bf \hat{\tilde P}} \right)$, ${\bf q}_\perp^2 \equiv  - \tilde q^2$
and
\bea
N_{\phi,R} &=& \frac{1}{R}  \int  \, d\tilde P \tilde P^3 |\tilde\phi({|{\bf \tilde
  P}|})|^2 
  \,,\nonumber\\
  N_{\phi,R,2} &=& \frac{R}{2}  \int  \, d\tilde P \tilde P |\tilde\phi({|{\bf\tilde
   P}|})|^2
  \,.
  \label{normaliz}
\eea
Unlike the electromagnetic case (see Refs.~\cite{Epelbaum:2022fjc} and \cite{Panteleeva:2022khw}), the dependence on the form of the profile function and the 
size of the packet remains in the normalization factors $N_{\phi,R}$ and $N_{\phi,R,2}$ of the local density distributions. 
Notice that for $R\to 0$ the first term in Eq.~(\ref{rhoint4G}) goes to infinity while the second term vanishes.
\medskip

Postponing the interpretation of the expression in Eq.~(\ref{rhoint4G}) to Sec.~\ref{interpretation} we now 
consider the ``static'' approximation. 
The local densities  of Refs.~\cite{Polyakov:2002wz,Polyakov:2002yz,Polyakov:2018zvc} in terms of the Fourier transforms of the form factors in the Breit frame  emerge by expanding 
the integrand in Eq.~(\ref{rhoint2G}) in powers of $1/m$ up to leading-order terms  before performing the integration \cite{Jaffe:2020ebz}
 and localizing the wave packet by taking $R\to 0$ limit.\footnote{
 Notice that the $R\to 0$ and $m\to \infty$ limits 
do not commute as pointed out in Ref.~\cite{Epelbaum:2022fjc}, and therefore the static expression does not emerge from Eq. \eqref{rhoint4G} by expanding it in powers of $1/m$.}
The resulting expressions after the $1/m$-expansion of the integrand have the form: 
\begin{align}
t^{00}_{\phi,{\rm naive}}( {\bf r}) & = m \int \frac{d^3 {P} \, d^3 q}{(2\pi)^3}\, \Theta_2\left(-{\bf q}^2 \right) 
                \phi\bigg({\bf P} -
\frac{\bf q}{2}\bigg) \, \phi^\star\bigg({\bf P} +\frac{\bf q}{2}\bigg)  \, e^{ - i {\bf q}\cdot {\bf  r}} , \nn
t^{0i}_{\phi,{\rm naive}}( {\bf r}) & = \int \frac{d^3 {P} \, d^3 q}{(2\pi)^3 }\, P^i \, \Theta_2\left( -{\bf q}^2 \right) 
                \phi\bigg({\bf P} -
\frac{\bf q}{2}\bigg) \, \phi^\star\bigg({\bf P} +\frac{\bf q}{2}\bigg)  \, e^{ - i {\bf q}\cdot {\bf  r}} , \nn
t^{ij}_{\phi,{\rm naive}}( {\bf r}) & = \frac{1}{2 m} \int \frac{d^3 {P} \, d^3 q}{(2\pi)^3}\, \bigl[ \left(  {\bf q}^2 \delta^{ij} -q^i q^j \right)\, \Theta_1\left(-{\bf q}^2 \right)  +2 P^i P^j \, \Theta_2\left( -{\bf q}^2  \right)   \bigr] 
                \phi\bigg({\bf P} -
\frac{\bf q}{2}\bigg) \, \phi^\star\bigg({\bf P} +\frac{\bf q}{2}\bigg)  \, e^{ - i {\bf q}\cdot {\bf  r}} .
\label{rhoint2NRG}
\end{align}
To localize the wave packet we take the $R \to 0$ limit again
by using the method of dimensional counting. The desired result is obtained by substituting ${\bf P}={\bf \tilde {
  P}}/R$, expanding the integrands in Eq.~(\ref{rhoint2NRG}) in powers of $R$
around $R=0$ and keeping explicitly only terms which are relevant, we obtain
\begin{eqnarray}
t^{00}_{{\rm naive}}( {\bf r}) 
& = & 
m \int \frac{d^3 {\tilde P} \, d^3 q}{(2\pi)^3}  \Theta_2\left(-{\bf q}^2 \right) \, |\tilde\phi({\bf{\tilde P}})|^2 e^{ - i {\bf q}\cdot {\bf  r}} +\text{Rest}  =  
  \int \frac{d^3 q}{(2\pi)^3} \, m\, \Theta_2\left(-{\bf q}^2 \right) 
                \, e^{ - i {\bf q}\cdot {\bf  r}} +\text{Rest}
\, , \nn
t^{0i}_{{\rm naive}}( {\bf r})  
& = & \int \frac{d^3 {\tilde P}\, d^3 q}{(2\pi)^3 }\, \frac{\tilde P^i}{R}  \, \Theta_2\left( -{\bf q}^2 \right) \, |\tilde\phi({\bf{\tilde
   P}})|^2  \, e^{ - i {\bf q}\cdot {\bf  r}} +\text{Rest} = 0 +\text{Rest} \,, \nn
t^{ij}_{{\rm naive}}( {\bf r}) 
& = & \frac{4\pi \, \delta^{ij}}{3 m R^2} \int d \tilde P \, \tilde P^4 \, 
 |\tilde\phi({\bf{\tilde  P}})|^2  
\int \frac{d^3 q}{(2\pi)^3} \, \Theta_2\left( -{\bf q}^2  \right) 
   e^{ - i {\bf q}\cdot {\bf  r}}   \nonumber\\
& + & \frac{1}{2m}\int \frac{d^3 q}{(2\pi)^3}\, 
\left( {\bf q}^2 \delta^{ij}-q^i q^j  \right) \Theta_1\left(-{\bf q}^2 \right) 
   e^{ - i {\bf q}\cdot {\bf  r}} +\text{Rest}
  \,,
\label{rhoint3NRG}
\end{eqnarray}
where  we used Eq.~(\ref{normN}) and the fact that for spherically symmetric packets integral over an odd function of ${\bf \tilde P}$ vanishes. 
The $t^{00}_{{\rm naive}}$ 
and the second term of $t^{ij}_{{\rm naive}}$ in Eq.~(\ref{rhoint3NRG})  coincide with the corresponding expressions of spatial densities obtained as the Fourier transforms of the gravitational form factors 
 in the Breit frame. Notice that both of these terms do not depend on the packet profile function and the size of the packet. 

\medskip

\subsection{Spin-1/2 particles}

The matrix elements of the EMT of a spin-1/2 system in one-particle eigenstates of the four-momentum operator  are parametrized in terms of three form factors as follows \cite{Polyakov:2018zvc}:
\begin{equation}
\langle p', s'| \hat T^{\mu\nu}({\bf x} ,0)| p,s \rangle = e^{ - i  {\bf q} \cdot {\bf x}} \bar u(p',s') \left[ A(q^2) \frac{P^\mu P^\nu}{m} + i J(q^2) \frac{P^\mu \sigma^{\nu\alpha} q_\alpha 
+ P^\nu \sigma^{\mu\alpha} q_\alpha}{2 m}+ D(q^2) \frac{q^\mu q^\nu-g^{\mu\nu} q^2}{4 m} \right]  u(p,s) \,,
\label{EMTdef}
\end{equation}  
where the momenta $q$ and $P$ are defined as in the previous section  and the Dirac spinors are normalized as $ \bar u(p,s') u(p,s) = 2 m\, \delta_{s's}$.
The matrix element of the EMT in localized states specified by Eq.~(\ref{statedefN2}) 
is written as 
\begin{align}
 t^{\mu\nu}_{\phi}(s',s,{\bf x} - {\bf X})  &\equiv \langle \Phi, {\bf X},s' | \hat
                                   T^{\mu\nu} ({\bf x}, 0 ) | \Phi, {\bf X},s
                                   \rangle 
                                    \nn &
  = \int \frac{d^3 {P} \, d^3 {q}}{(2\pi)^3 \sqrt{4 E
    E'}}\, 
\, \bar u(p',s') \left[ A\left((E-E')^2- {\bf q}^2\right) \frac{P^\mu P^\nu}{m} +  i J\left((E-E')^2- {\bf q}^2\right) \frac{P^\mu \sigma^{\nu\alpha} q_\alpha 
+ P^\nu \sigma^{\mu\alpha} q_\alpha}{2 m}  \right. \nn
& \left. + D\left((E-E')^2- {\bf q}^2\right) \frac{q^\mu q^\nu-g^{\mu\nu} q^2}{4 m}
 \right] u(p,s)    \,  \phi\bigg({\bf P} -
\frac{\bf q}{2}\bigg) \, \phi^\star\bigg({\bf P} +\frac{\bf q}{2}\bigg)  \, e^{ - i {\bf q}\cdot ({\bf  x} - {\bf  X})} .
\label{rhoint2}
\end{align}  
Taking the $R\to 0$ limit in Eq.~(\ref{rhoint2}) by using the method of dimensional counting in a complete analogy to the case of spin-0 systems, and 
assuming that form factors  $A(q^2)$, $J(q^2)$ and $D(q^2)$ decay for large $q^2$ as $1/q^2$, $1/(q^2)^{3/2}$ and $1/q^4$, or faster, respectively,\footnote{This condition 
 is actually satisfied by the large-$q^2$ behavior of the gravitational form factors of the proton
 obtained in perturbative QCD in Refs.~\cite{Tanaka:2018wea,Tong:2021ctu,Tong:2022zax}, 
in our notation corresponding to $A(q^2)\sim J(q^2)\sim 1/q^4$ and $D(q^2)\sim 1/q^6$, modulo logarithms.}
we obtain for the operators in spin space
\begin{align}
t^{\mu\nu}_{\phi}({\bf r})  & =  N_{\phi,R}
 \int \frac{d^2 \hat{\tilde P} \, d^3 {q}}{(2\pi)^3 } 
 \left[ \frac{i}{2 m} \left(\hat {\tilde P}^{\mu}({\boldsymbol \sigma}_\perp\times{\bf q})^\nu +\hat {\tilde P}^{\nu}({\boldsymbol \sigma}_\perp\times{\bf q})^\mu+  
{\bf \hat{\tilde P}}\cdot ({\boldsymbol \sigma}_\perp\times{\bf q})(\delta^{\mu 0}\hat {\tilde P}^{\nu}+\delta^{\nu 0}\hat {\tilde P}^\mu) 
\right)
J\left( - {\bf q}_\perp^2\right)\right.\nn 
&\left.+\hat {\tilde P}^\mu \hat {\tilde P}^\nu  A\left( - {\bf q}_\perp^2\right) 
\right] e^{ - i {\bf q}\cdot {\bf  r}} 
  + \frac{1}{2} \, N_{\phi,R,2} \,  \int \frac{d^2 \hat{\tilde P} \, d^3 {q}}{(2\pi)^3 } \, 
\, \left( \tilde q^\mu \tilde q^\nu + g^{\mu\nu} {\bf q}_\perp^2 \right) \, D\left(  - {\bf q}_\perp^2 \right)   
 \, e^{ - i {\bf q}\cdot {\bf  r}}+\text{Rest} \,,
\label{rhoint2NRGR}
\end{align}
where ${\bf r}={\bf x}-{\bf X}$ and $({\boldsymbol
    \sigma}_\perp\times{\bf q})^0=0$. Again, we wrote
  explicitly only those terms which contribute to the final expressions for the densities in Sec.~\ref{interpretation}.

Postponing again the discussion of the obtained result to the Sec.~\ref{interpretation}, we consider next the static approximation. 
Expanding the integrand in Eq.~(\ref{rhoint2}) in powers of $1/m$ up to leading-order terms we obtain  for the operators in spin space
\begin{align}
t^{00}_{\phi,{\rm naive}}( {\bf r})  &  = m \int \frac{d^3 {P} \, d^3 {q}}{(2\pi)^3 } \, 
\,   A\left( - {\bf q}^2\right) \phi\bigg({\bf P} -
\frac{\bf q}{2}\bigg) \, \phi^\star\bigg({\bf P} +\frac{\bf q}{2}\bigg)  \, e^{ - i {\bf q}\cdot {\bf  r}}
, \nn 
t^{0i}_{\phi,{\rm naive}}({\bf r})  & =  \int \frac{d^3 {P} \, d^3 {q}}{(2\pi)^3} \,  \left[ 
A\left( - {\bf q}^2\right) P^i   - \frac{i}{2} \, \epsilon^{ijk} q^j \sigma^k \,  J\left( - {\bf q}^2\right)  
 \right] 
                \phi\bigg({\bf P} -
\frac{\bf q}{2}\bigg) \, \phi^\star\bigg({\bf P} +\frac{\bf q}{2}\bigg)  \, e^{ - i {\bf q}\cdot {\bf  r}} , \nn 
t^{ij}_{\phi,{\rm naive}}({\bf r})  & = \frac{1}{m} \int \frac{d^3 {P} \, d^3 {q}}{(2\pi)^3}\,
\left[ 
A\left( - {\bf q}^2\right) \, P^i P^j + \frac{1}{4} \,
D\left( - {\bf q}^2\right) \, \left( - {\bf q}^2 \delta^{ij} +q^i q^j\right)  \right. \nn
& \left. - \frac{i}{2} \, \left( \epsilon^{ilk} q^l P^j + \epsilon^{jlk} q^l P^i  \right) \sigma^k \, J\left( - {\bf q}^2\right) 
\right] 
                \phi\bigg({\bf P} -
\frac{\bf q}{2}\bigg) \, \phi^\star\bigg({\bf P} +\frac{\bf q}{2}\bigg)  \, e^{ - i {\bf q}\cdot {\bf  r}} .
\label{rhoint2NR}
\end{align}
To localize the wave packet by taking the limit $R \to 0$
we use the method
of dimensional counting and obtain the following expressions for the ``naive"  
spatial densities:
\begin{eqnarray}
t^{00}_{{\rm naive}}({\bf r}) 
&  = & m \int \frac{ d^3 {q}}{(2\pi)^3 } 
\,  A\left( - {\bf q}^2\right)  e^{ - i {\bf q}\cdot {\bf  r}} +\text{Rest}
, \nn 
t^{0i}_{{\rm naive}}( {\bf r})  
& = & - \frac{i}{2} \epsilon^{ijk}   \sigma^k   \int \frac{d^3 {q}}{(2\pi)^3 } q^j   J\left( - {\bf q}^2\right) 
 e^{ - i {\bf q}\cdot {\bf  r}}  +\text{Rest} , \nn 
t^{ij}_{\phi,{\rm naive}}({\bf r}) 
& = & \frac{1}{R^2}  \int  \, d\tilde P \tilde P^4 |\tilde\phi({\bf{\tilde
  P}})|^2 \, \frac{4\pi \, \delta^{ij} }{3m} \int \frac{d^3 {q}}{(2\pi)^3}\,
 \,A\left( - {\bf q}^2\right) 
 \, e^{ - i {\bf q}\cdot {\bf  r}}  
 \nn  
 & + & \frac{1 }{4m} \int \frac{d^3 {q}}{(2\pi)^3}
D\left( - {\bf q}^2\right) \, \left( - {\bf q}^2 \delta^{ij} +q^i q^j\right)  
\, e^{ - i {\bf q}\cdot {\bf  r}} +\text{Rest} \,,
\label{rhoint3RGR}
\end{eqnarray}
where we made use of Eq.~(\ref{normN}) and the fact that the integral over ${\bf \tilde P}$ of an odd function of this variable vanishes. 
Analogously to the case of a spin-0 system, the $t^{00}_{{\rm naive}}$, $t^{0i}_{{\rm naive}}$ and the second term of $t^{ij}_{\phi,{\rm naive}}$ in Eq.~(\ref{rhoint3RGR})  coincide with the corresponding expressions 
of spatial densities obtained as the Fourier transforms of the gravitational form factors 
 in the Breit frame. 

Defining the mean squared energy radius of a system via 
\begin{equation}
\langle r^2\rangle = \frac{\int d^3 r \, t^{00}({\bf r}) r^2 }{ \int d^3 r \, t^{00}({\bf r})} \,,
\label{defrad}
\end{equation}
for a spin-1/2 system in the static approximation of Eq.~(\ref{rhoint3RGR}) we obtain $\langle r^2\rangle_{{\rm naive}} = 6 A'(0)$, while Eq.~(\ref{rhoint2NRGR}) leads to 
$\langle r^2\rangle = 4 A'(0)$. Analogous result holds also for spin-0 systems. The same smaller mean squared radius, compared to the Breit-frame result, has been obtained previously 
in Ref.~\cite{Freese:2021czn} using the two-dimensional formalism utilizing light-front coordinates.   

\section{Gravitational densities in moving frames}
\label{MFrames}

\subsection{Spin-0}

To generalize the results of the previous section  to moving frames
we replace $\phi ({\bf p} )$ with its boosted expression $\phi_{\bf v} ({\bf p} )$, 
given in terms of the spherically symmetric ZAMF quantity as follows \cite{Hoffmann:2018edo} 
\be
\label{temp01}
\phi_{\bf v} ({\bf p})= \sqrt{\gamma \Big( 1-\frac{{\bf v} \cdot {\bf
      p}}{E} \Big)} \, \phi \big[ {\bf p}_\perp + \gamma ( {\bf
  p}_\parallel - {\bf v} E )\big]\,,
\ee
where ${\bf v}$ is the boost velocity, $\gamma = (1 - v^2)^{-1/2}$, ${\bf p}_\parallel = ({\bf p} \cdot
{\bf\hat {v}}) {\bf\hat  v}$, ${\bf p}_\perp = {\bf p} - {\bf p}_\parallel$ and
$E = \sqrt{m^2 + {\bf p}^2}$. 

In exact analogy to the case of the ZAMF,  we obtain for a spin-0 system 
in the $R \to 0$ limit using the method of dimensional counting:
\begin{equation}
t^{\mu\nu}_{\phi,  {\bf v}} ( {\bf r})\!\! =\!\! \int\! \frac{d^3 \tilde {P} \, d^3
  {q}}{(2\pi)^3} \,
\gamma \left(\!1\!  -\! {\bf v} \cdot {\bf \hat{\tilde P}}\! \right)\!
 \Biggl\{\! \frac{ \tilde P\,  \hat{\tilde P}^\mu \hat{\tilde P}^\nu}{  R}  \Theta_2 \left[ \tilde q^2
\right]  
\!+\! \frac{R}{2 \tilde P} \left[ \tilde q^2
g^{\mu\nu}\! -\! \tilde q^\mu \tilde q^\nu  \right]\!
\Theta_1\! \left[ \tilde q^2
\right]\!   \Biggl\} \!
      \, \Big|\tilde\phi \left[
  {\bf\tilde P}_\perp\! +\! \gamma (\bf{  \tilde  P}_\parallel\! -\! {\bf v}
      \tilde {P} ) \right] \!\Big|^2 \! e^{- i {\bf q}\cdot {\bf  r}}  +\text{Rest},
\label{rhoint3boosted}
\end{equation}
where ${\bf \hat {\tilde  P}}$, $\hat{\tilde P}^\mu$  and $\tilde q^\mu$ are defined as in the previous section.
 Next, we change the integration variable ${\bf\tilde P} \to{\bf \tilde P}' =
 {{\bf\hat  v}} \times \big( {\bf\tilde P} \times  {\bf\hat v}\big)   +
  \gamma \big({\bf\tilde  P}  \cdot  {\bf\hat    v}   -   v \tilde
  P\big)    {\bf\hat v}$, denote ${\bf\hat {m}} \equiv 
{\bf\hat{\tilde P}'}$ and introduce a vector-valued function
\begin{equation}
  {\bf n} \big({\bf v},  {\bf\hat m} \big) = {\bf\hat  v} \times \big( {\bf\hat  m} \times  {\bf\hat  v}\big)
+ \gamma \big( {\bf\hat  m} \cdot  {\bf\hat  v}  + v )  {\bf\hat v} \,.
\end{equation}
Taking into account that $\tilde{\bf P} =
 {\bf\hat  v} \times \big( {\bf\tilde P}' \times  {\bf\hat  v}\big)   +
  \gamma \big({\bf\tilde  P}'  \cdot  {\bf\hat  v}   +   v \tilde
  P'\big)    {\bf\hat  v}$, it follows that  $ {\bf\hat { n}} =  {\bf\hat {\tilde {P}}} $. 
The Jacobian of the change of variables ${\bf\tilde P} \to {\bf\tilde P}'$
cancels the first factor in the integrand in Eq.~(\ref{rhoint3boosted}), and we finally obtain
\begin{eqnarray}
  \label{rhoint3boosted2}
t^{00}_{\phi,  {\bf v}} ( {\bf r})\!\!\! &=&\!\!\!\!\! \int \frac{d {\bf\hat m} \, d \tilde {P}' \tilde {P}'^2 \, d^3
  {q}}{(2\pi)^3}
                     \; \big|\tilde\phi \big(
{\bf\tilde  P} ' \big) \big|^2  \, e^{ - i {\bf q}\cdot {\bf  r}} 
              \Biggl\{  \frac{\gamma ( \tilde {P}' + {
                             v}  \tilde {P}_\parallel ') }{R} \,                   \Theta_2\left[ ({\bf \hat n \cdot q})^2 - {\bf q}^2 \right] 
              -  \frac{  {\bf q}^2 R}{2 \gamma ( \tilde {P}' + {
                             v}  \tilde {P}_\parallel ')}         \,       \Theta_1\left[  
                             ({\bf \hat n \cdot q})^2 - {\bf q}^2 \right] \Biggr\} +\text{Rest} \,,
                \nonumber \\
 t^{0i}_{\phi,  {\bf v}} ( {\bf r})\!\!\! &=&\!\!\!\!\! \int \frac{d{\bf \hat m} \, d \tilde {P}' \tilde {P}'^2 \, d^3
  {q}}{(2\pi)^3} 
                                   \;  \big|\tilde\phi \big(
{\bf\tilde  P} ' \big) \big|^2  \, e^{ - i {\bf q}\cdot {\bf  r}}  \Biggl\{  \frac{ \gamma ( \tilde {P}' + {
                             v}  \tilde {P}_\parallel ')  
                              {\hat n}^i }{R} 
              \,                    \Theta_2\left[ 
                            ({\bf \hat n \cdot q})^2\! -\! {\bf q}^2 \right] 
           \!\!    -\!\! \frac{R}{2}  
                                \frac{  {\bf \hat n \cdot q} }{\gamma ( \tilde {P}' + {
                             v}  \tilde {P}_\parallel ') }   \,  q^i 
                                  \Theta_1\left[  
                                  ({\bf \hat n \cdot q})^2 \!-\! {\bf q}^2 \right]\!\! \Biggr\} +\text{Rest}
                \,,
                \nonumber \\
t^{ij}_{\phi,  {\bf v}} ( {\bf r})\!\!\! &=&\!\!\!\!\! \int \frac{d {\bf\hat m} \, d \tilde {P}' \tilde {P}'^2\, d^3
  {q}}{(2\pi)^3}
                                   \;  \big|\tilde\phi \big(
{\bf\tilde  P} ' \big) \big|^2  
 e^{ - i {\bf q}\cdot {\bf  r}}  \Biggl\{  \frac{ \gamma ( \tilde {P}' + {
                             v}  \tilde {P}_\parallel ')
 }{ R} \, {\hat n}^i  
                                   {\hat n}^j 
                      \Theta_2\left[ ({\bf \hat n \cdot q})^2 - {\bf q}^2 \right] \nn
               &- &  \frac{R}{2}  
                                   \;  \frac{ 1}{ \gamma ( \tilde {P}' + {
                             v}  \tilde {P}_\parallel ')}\, 
                             \left[ \left(  
                            ({\bf \hat n \cdot q})^2 - {\bf q}^2 \right)  \delta^{ij} + q^i q^j  \right] 
                \Theta_1\left ({\bf \hat n \cdot q})^2 - {\bf q}^2 \right] \Biggr\} +\text{Rest}\,.             
\end{eqnarray}
Using the spherical symmetry of $\tilde\phi \big(
{\bf\tilde {P}} ' \big)$, the integration over $\tilde P'$ factorizes out in Eq.~(\ref{rhoint3boosted2}).
To carry out the remaining angular integration over $ {\bf\hat  m}$ in
spherical coordinates we align the $z$- and $x$-axes along the ${\bf v}$
and ${\bf q}_\perp$ directions, respectively, denote $\eta =
\cos \theta $ and obtain
\be
\label{RhoBoostedCoord1} 
t^{\mu\nu}_{\bf v} ({\bf r}) = N_{\phi, R} \int \frac{d^3 {q}}{(2\pi)^3}\, \bar t^{\,\mu\nu} \left(
  q_\parallel,   {q}_\perp \right)  \, e^{ - i {\bf q}\cdot {\bf  r}} + N_{\phi, R,2} \int \frac{d^3 {q}}{(2\pi)^3}\, \bar t_2^{\,\mu\nu} \left(
  q_\parallel,   {q}_\perp \right)  \, e^{ - i {\bf q}\cdot {\bf  r}}+\text{Rest}\,,
\ee
with $q_\parallel \equiv {\bf\hat v} \cdot {\bf q}$, $q_\perp \equiv | {\bf q}_\perp |$ and 
\begin{align}
\bar t^{\, \mu\nu} (
  q_\parallel,   {q}_\perp ) & = \int_{-1}^{+1}
d\eta \int_0^{2 \pi} d\phi \,
\frac{\Omega^\mu   \Omega^\nu}{\gamma(1+ v \eta )}
                             \Theta_2\left[ 
 {\bar q}^2  \right] \,, \nn
\bar t_2^{\, \mu\nu } (
  q_\parallel,   {q}_\perp ) & = - \int_{-1}^{+1}
d\eta \int_0^{2 \pi} d\phi \,  \frac{ 1}{ \gamma ( 1 + {
                             v} \eta)}\, 
                             \left[  \bar q^\mu \bar q^\nu -  {\bar q} ^2   g^{\mu\nu} \right] 
                              \Theta_1\left[ {\bar q}^2  \right] \,,                                
\label{RhoBoostedCoord2}
\end{align}
where $\Omega^\mu = (\gamma( 1+ v \eta) , {\bf  \hat \omega_\perp  + \gamma ( \hat \omega_\parallel  + {\bf v}) } ) $,
$\bar q^\mu = ( \big[\sqrt{1-\eta^2} \cos \phi \, q_\perp + \gamma (\eta + v) q_\parallel \big]/ (\gamma (1 + v  \eta)) , {\bf q})$ and ${ \hat\omega} =(\sqrt{1-\eta^2} \cos\phi , \sqrt{1-\eta^2} \sin\phi , \eta) $.

\medskip

In the infinite momentum frame (IMF) with  $v \to 1$, $\gamma \to \infty$,  we obtain 
\begin{align}
\bar t^{\, \mu\nu} (
  q_\parallel,   {q}_\perp ) & 
  =   4 \pi \, \gamma \, \hat v^{\mu} \hat v^{\nu}\,
 \Theta_2\left[  - {\bf q}_\perp^2 
 \right] \,,\nn
\bar t_2^{\, \mu\nu} (
  q_\parallel,   {q}_\perp ) & 
  =   - \frac{2 \pi}{\gamma}  \,  \alpha \, \left[   {\bf q}_\perp^2     g^{\mu\nu} +  q_v^\mu q_v^\nu \right]  
 \Theta_1\left[  - {\bf q}_\perp^2 
 \right] 
\,,
\label{RhoBoostedCoord2GR}
\end{align}
where $\hat v^\mu =(1,{\bf \hat v})$, $q_v^\mu =(q_\parallel , {\bf q})$ and
\be
\alpha =\lim_{v\to1} \int_{-1}^{+1} \frac{d\eta}{1+v\eta}\,.
\label{defdiv}
\ee
Notice that $\alpha/\gamma \sim \sqrt{1-v} \ln (1-v)$ when the $v\to 1$ limit is taken.
Substituting these expressions into Eq.~(\ref{RhoBoostedCoord1}) for the IMF we obtain:
\bea
\label{RhoBoostedCoord1RPPR} 
t^{00}_{\rm IMF} ({\bf r}) &=& N_{\infty}  \, \delta(r_\parallel)   
\tilde \Theta_2\left[   {\bf r}_\perp  \right] 
  + N_{0} 
 \left[ \delta(r_\parallel) \, \frac{\partial^2  \tilde \Theta_1\left[   {\bf r}_\perp 
 \right] }{\partial {\bf r}_\perp^k \partial {\bf r}_\perp^k }  
  + \frac{\partial^2 \delta(r_\parallel)}{\partial {\bf r}_\parallel^k \partial {\bf r}_\parallel^k }   
\tilde \Theta_1\left[  {\bf r}_\perp 
 \right]  \right]  +\text{Rest}
\,, 
 \nonumber\\
 t^{0i}_{\rm IMF} ({\bf r}) &=& N_{\infty}  \, \delta(r_\parallel) 
 \hat v^i \tilde \Theta_2\left[   {\bf r}_\perp  \right] 
 + N_{0} \, \hat v^k \left( \frac{\partial^2 \delta(r_\parallel) }{\partial {\bf r}_\parallel^k \partial {\bf r}_\parallel^i }  
 \tilde \Theta_1\left[   {\bf r}_\perp  \right] + \frac{\partial \delta(r_\parallel) }{\partial {\bf r}_\parallel^k }  
\frac{\partial \tilde \Theta_1\left[   {\bf r}_\perp  \right] }{\partial {\bf r}_\perp^i } \right) +\text{Rest}\,,
\nonumber\\
 t^{ij}_{\rm IMF} ({\bf r}) &=& N_{\infty}  \, \delta(r_\parallel) 
 \hat v^i \hat v^j \tilde \Theta_2\left[   {\bf r}_\perp  \right] 
 + N_{0}\biggl( - \delta(r_\parallel) \, \frac{\partial^2  \tilde \Theta_1\left[   {\bf r}_\perp 
 \right] }{\partial {\bf r}_\perp^k \partial {\bf r}_\perp^k}\,\delta^{ij} 
 + \frac{\partial^2 \delta(r_\parallel) }{\partial {\bf r}_\parallel^i \partial {\bf r}_\parallel^j }  \tilde \Theta_1\left[   {\bf r}_\perp  \right] 
 + \frac{\partial \delta(r_\parallel) }{\partial {\bf r}_\parallel^i }  \frac{\partial \tilde \Theta_1\left[   {\bf r}_\perp  \right] }{\partial {\bf r}_\perp^j } \nonumber\\
 & + & 
 \frac{\partial \delta(r_\parallel) }{\partial {\bf r}_\parallel^j }  \frac{\partial \tilde \Theta_1\left[   {\bf r}_\perp  \right] }{\partial {\bf r}_\perp^i } 
 +  \delta(r_\parallel)   \frac{\partial \tilde \Theta_1\left[   {\bf r}_\perp  \right] }{\partial {\bf r}_\perp^i \partial {\bf r}_\perp^j } 
 \biggr)+\text{Rest} \,,
\eea
where $ N_{0} = \frac{2 \pi  \,  \alpha }{\gamma}   N_{\phi, R,2}$, $N_{\infty} = 4 \pi  \gamma  N_{\phi, R}  $. Furthermore  the Fourier transforms of the form factors are given by
\begin{eqnarray}
\tilde \Theta_i\left[  {\bf r}_\perp 
 \right]  =  \int \frac{d^2 {q}_\perp}{(2\pi)^2}\, \Theta_i\left[  - {\bf q}_\perp^2 
 \right] e^{ - i {\bf q}_\perp \cdot {\bf  r}_\perp}  .
\label{defFFr}
\end{eqnarray}

\subsection{Spin-1/2}

For a spin-1/2 system in a moving frame, we consider the state specified by the following wave packet
\begin{eqnarray}
|\Phi, {\bf X},s \rangle _{\bf v} & = & \int \frac{d^3 {p}}{\sqrt{2 E (2\pi)^3}}  \, \sqrt{\gamma \Big( 1-\frac{{\bf v} \cdot {\bf
      p}}{E} \Big)} \, \phi \big[ \Lambda_{\bf v}^{-1} {\bf p}
\big]
\, e^{-i {\bf p}\cdot{\bf X}} 
\sum_{s_1}D_{s_1s}\Big[ W\Big( \Lambda_{\bf v}, 
    \frac{ \Lambda_{\bf v}^{-1}  {\bf p}  }{m} \Big) 
 \Big]  |p ,s_1 \rangle,  
\label{statedefN2Moving}
\end{eqnarray}
where $\gamma = (1 - v^2)^{-1/2}$,
$E = \sqrt{m^2 + {\bf p}^2}$,  $\Lambda_{\bf v}^{-1} {\bf p} = 
     {\bf\hat { v}} \times \big(  {\bf p} \times  {\bf\hat { v}}\big)
       +  \gamma   \big( {\bf   p} \cdot  {\bf\hat  { v}}  - v E\big)  {\bf\hat { v}} $ with $\Lambda_{\bf v}$ denoting the Lorentz boost from the ZAMF to the moving frame.
   The $D_{s_1s}\left[ W\right]$ matrices build the spin-1/2 representation of the Wigner rotations 
     \cite{Weinberg:1995mt}.

For the matrix elements of the EMT of a spin-1/2 system in a moving frame we obtain  the following expressions in the $R\to 0$ limit using the method of dimensional counting:
\begin{eqnarray}
t^{00}_{\phi,  {\bf v}}({\bf r}) & = & \int \frac{d^3 {\tilde P} \, d^3 {q}}{(2\pi)^3} 
\, \gamma \left(1  - {\bf v} \cdot {\bf \hat{ {\tilde P}}}\right) 
\left\{ \frac{\tilde P}{R} 
\left( 1 - \frac{i}{2 m }\, \epsilon^{ijk} \sigma^k q^i {\hat{\tilde P}}^j\right)
 A \left[ \tilde q^2
 \right]   
+ \frac{\tilde P q^i}{2 R m} 
\, \left[ \Sigma^i_{\bf v,\hat {m}} ,  {\bf \hat{\tilde P}} \cdot {\bm\sigma} \right]_{-}  J \left[ \tilde q^2 \right]  
 \right. \nn  & + &  \left.
  \frac{R \,  {\bf q}^2}{4 \tilde P} \, \left( 1 - \frac{i}{2 m}\, \epsilon^{l n k} \sigma^k q^l {\hat{\tilde P}}^n \right) D \left[ \tilde q^2
 \right] 
  \right\} 
 \Big|\tilde\phi \bigl(  {\bf\tilde P}' \bigr) \Big|^2   \, e^{ - i {\bf q}\cdot {\bf  r}}+\text{Rest}\,, \nn
t^{0i}_{\phi,  {\bf v}}({\bf r}) & = & \int \frac{d^3 {\tilde P} \, d^3 {q}}{(2\pi)^3}\, \gamma \left(1  - {\bf v} \cdot {\bf \hat{ {\tilde P}}}\right) 
 \Biggl\{ 
\tilde P\, \frac{{\hat{\tilde P}}^i}{R } \left( 1 - \frac{i}{2 m}\, \epsilon^{l jk} \sigma^k q^l {\hat{\tilde P}}^j \right) A\left[ \tilde q^2
\right]  
  \nonumber\\ & 
- & \frac{i \tilde P}{4 R \,m} \biggl[  
 i \, {\hat{\tilde P}}^i q^j   \left[ \Sigma^j_{\bf v,\hat{m}} ,  {\bf \hat{\tilde P}} \cdot {\bm\sigma} \right]_{-}
 +  i \, {\bf \hat{\tilde P}}  \cdot {\bf q}  \left[ \Sigma^i_{\bf v,\hat{m}} ,  {\bf \hat{\tilde P}} \cdot {\bm\sigma} \right]_{-}-\epsilon^{ijk} q^j {\bf \hat{\tilde P}} \cdot {\bm\sigma} \left[ \Sigma^k_{\bf v,\hat{m}} ,  {\bf \hat{\tilde P}} \cdot {\bm\sigma} \right]_{-}
\biggr]   
J \left[ \tilde q^2
\right] 
\nn
&+ & \frac{R\, {\bf \hat{\tilde P}}  \cdot {\bf q}  \, q^i}{4 \tilde P}\, \left( 1 - \frac{i}{2 m}\, \epsilon^{l n k} \sigma^k q^l {\hat{\tilde P}}^n \right)  \, D\left[ \tilde q^2
\right] 
 \Biggr\}                                  
      \, \Big|\tilde\phi \bigl(  {\bf\tilde  P}' \bigr) \Big|^2   \, e^{ - i {\bf q}\cdot {\bf  r}}+\text{Rest}\,,
      \nn
t^{ij}_{\phi,  {\bf v}}({\bf r}) & = & \int \frac{d^3 {\tilde P} \, d^3 {q}}{(2\pi)^3}\, \gamma \left(1  - {\bf v} \cdot {\bf \hat{ {\tilde P}}}\right)   
\Biggl\{ \frac{\tilde P \, {\hat{\tilde P}}^i {\hat{\tilde P}}^j }{R}\, \left( 1 - \frac{i}{2 m}\, \epsilon^{l n k} \sigma^k q^l {\hat{\tilde P}}^n \right) A\left[ \tilde q^2 \right] 
\nonumber\\ 
& 
- & \frac{i \tilde P}{4 R \,m} \biggl[ 
 i \, {\bf \hat{\tilde P}}  \cdot {\bf q}  \left( {\hat{\tilde P}}^i   \left[ \Sigma^j_{\bf v,\hat{m}} ,  {\bf \hat{\tilde P}} \cdot {\bm\sigma} \right]_{-}
 +  {\hat{\tilde P}}^j  \left[ \Sigma^i_{\bf v,\hat{m}} ,  {\bf \hat{\tilde P}} \cdot {\bm\sigma} \right]_{-} \right)\nn
&-& \left( \epsilon^{jkl} {\hat{\tilde P}}^i q^k + \epsilon^{ikl} {\hat{\tilde P}}^j q^k \right) {\bf \hat{\tilde P}} \cdot {\bm\sigma}\left[ \Sigma^l_{\bf v,\hat{m}} ,  {\bf \hat{\tilde P}} \cdot {\bm\sigma} \right]_{-}
\biggr]   
J \left[ \tilde q^2
\right] 
\nn
& + &  \frac{R (\tilde q^i \tilde q^j + \delta^{ij} \tilde q^2)}{4 \tilde P } \left( 1 - \frac{i}{2 m}\, \epsilon^{l n k} \sigma^k q^l {\hat{\tilde P}}^n \right)  \, 
D\left[ \tilde q^2 \right] 
 \Biggr\}                              
      \, \Big|\tilde\phi \bigl( {\bf \tilde  P}' \bigr) \Big|^2  \, e^{ - i {\bf q}\cdot {\bf  r}}+\text{Rest}\,,
\label{rhoint3boostedGR}
\end{eqnarray}
where $[A,B]_{-}=A B-B A$, $ {\bf\tilde P}' =
 {\bf\hat { v}} \times \big( {\bf\tilde  P} \times  {\bf\hat { v}}\big)   +
  \gamma \big({\bf\tilde   P}  \cdot  {\bf\hat  {   v}}   -   v \tilde
  P\big)    {\bf\hat { v}}$ 
and the unit vector ${\bf\hat{m}}$ is defined as ${\bf\hat{m}}\equiv {\bf\hat {\tilde P}}'$. Further, the $\Sigma^i_{\bf v,\hat{m}} $ refers to the Wigner-rotated spin operator 
\begin{equation}
\Sigma^i_{\bf v,\hat{m}} = D^\dagger \left[ W(\Lambda_{\bf v, \hat m}) \right] \sigma^i D\left[ W(\Lambda_{\bf v, \hat m}) \right],
\label{defSigma}
\end{equation}
with
\begin{equation}
D\left[ W(\Lambda_{\bf v, \hat m}) \right] = \lim_{R\to 0} D\left[ W\Big( \Lambda_{\bf v}, 
    \frac{  \Lambda^{-1}_{\bf v} \big( {\bf \tilde  P}/R\big)   }{m} \Big) \right].
       \label{defW}
       \end{equation}
  Next, we change the integration variables ${\bf\tilde P} \to {\bf\tilde P}' $  to obtain
\bea
t^{00}_{\phi,  {\bf v}}({\bf r})\!\!\! & =\!\!\! & \frac{1}{R} \int \frac{d{\bf\hat{m}} \, d {\tilde P'} \tilde {P}'^2\, d^3 {q}}{(2\pi)^3}  \; \big|\tilde\phi \big(
{\bf\tilde P} ' \big) \big|^2 
\, \gamma ( \tilde {P}' + {
                             v}  \tilde {P}_\parallel ') \left\{
\left( 1 - \frac{i}{2 m}\, \epsilon^{ijk} \sigma^k q^i   
{\hat n}^j \right)
A\left[  ({\bf \hat n \cdot q})^2 - {\bf q}^2 
 \right]  \right. 
 \nn  & + & 
 \left.
 \frac{1}{2 m} \, q^i 
 \left[ \Sigma^i_{\bf v,\hat{m}} ,  {\bf \hat{n}} \cdot {\bm\sigma} \right]_{-}  
J \left[ ({\bf \hat n \cdot q})^2 - {\bf q}^2 \right]  
   \right\}                                
       \, e^{ - i {\bf q}\cdot {\bf  r}} \nn
  & + &    \frac{R}{2} \int \frac{d{\bf\hat{ m}} \, d {\tilde P'} \tilde {P}'^2\, d^3 {q}}{(2\pi)^3}  \;   \big|\tilde\phi \big(
{\bf\tilde P} ' \big) \big|^2  \,
\frac{{\bf q}^2}{2 \gamma ( \tilde {P}' + {
                             v}  \tilde {P}_\parallel ')}  \left( 1 - \frac{i \, \epsilon^{l n k} \sigma^k q^l  
                             {\hat n}^n}{2 \,m  
                             }\right)   
D\left[  ({\bf \hat n \cdot q})^2 - {\bf q}^2
 \right]  e^{ - i {\bf q}\cdot {\bf  r}}+\text{Rest}
       \,, \nn
t^{0i}_{\phi,  {\bf v}}({\bf r}) \!\!\! & =\!\!\! & \frac{1}{R} \int \frac{d{\bf\hat{ m}} \, d {\tilde P'} \tilde {P}'^2\, d^3 {q}}{(2\pi)^3}\, 
 \; \big|\tilde\phi \big(
{\bf\tilde  P} ' \big) \big|^2  \,
\, \gamma ( \tilde {P}' + {
                             v}  \tilde {P}_\parallel ') \,
  \biggl\{  
  {\hat n}^i  \, \left( 1 - \frac{i \, \epsilon^{l jk} \sigma^k q^l 
 {\hat n}^j}{2 \,m 
                             }  \right) 
                             A\left[  ({\bf \hat n \cdot q})^2 - {\bf q}^2 \right]   
 \nonumber\\ 
 & 
- & \frac{i}{4 m} \biggl[ 
  i \, { \hat{n}^i} q^j   
 \left[ \Sigma^i_{\bf v,\hat{m}} ,  {\bf \hat{n}} \cdot {\bm\sigma} \right]_{-} 
 +  i \, {\bf \hat{n}}\cdot{\bf q}   \left[ \Sigma^i_{\bf v,\hat{m}} ,  {\bf \hat{n}} \cdot {\bm\sigma} \right]_{-} -\epsilon^{ijk} q^j {\bf \hat{n}} \cdot {\bm\sigma}\left[ \Sigma^i_{\bf v,\hat{m}} ,  {\bf \hat{n}} \cdot {\bm\sigma} \right]_{-} 
\biggr]   
J \left[ ({\bf \hat n \cdot q})^2 - {\bf q}^2 \right] 
 \biggr\}  \, e^{ - i {\bf q}\cdot {\bf  r}}  \nn
  & + &    \frac{R}{2} \int \frac{d{\bf\hat{ m}} \, d {\tilde P'} \tilde {P}'^2 \, d^3 {q}}{(2\pi)^3}  \;   \big|\tilde\phi \big(
{\bf\tilde P} ' \big) \big|^2  \,
\frac{ {\bf \hat n \cdot q} \, {q}^i}{2 \gamma ( \tilde {P}' + {
                             v}  \tilde {P}_\parallel ')}  \left( 1 - \frac{i \, \epsilon^{l n k} \sigma^k q^l 
                              {\hat n}^n}{2 \,m }\right)  
D\left[ ({\bf \hat n \cdot q})^2 - {\bf q}^2 \right]  \, e^{ - i {\bf q}\cdot {\bf  r}} +\text{Rest},
   \nn
t^{ij}_{\phi,  {\bf v}}({\bf r}) \!\!\! & =\!\!\! & \frac{1}{R} \int \frac{d{\bf\hat{ m}} \, d {\tilde P' \tilde {P}'^2} \, d^3 {q}}{(2\pi)^3}\, 
 \; \big|\tilde\phi \big(
{\bf\tilde P} ' \big) \big|^2  \,
\gamma \left( \tilde {P}' + { v}  \tilde {P}_\parallel ' \right)  \biggl\{ 
{\hat n}^i   {\hat n}^j 
                             \left( 1 - \frac{i}{2 m}\, \epsilon^{l n k} \sigma^k q^l 
                             { \hat n}^n 
                             \right) A\left[  ({\bf \hat n \cdot q})^2 - {\bf q}^2 \right]    \nonumber\\ 
  & 
- & \frac{i }{4 m} \biggl[
 i \, {\bf \hat{n}}\cdot{\bf q} \left( {\hat n}^i    \left[ \Sigma^j_{\bf v,\hat{m}} ,  {\bf \hat{n}} \cdot {\bm\sigma} \right]_{-} 
   +  {\hat n}^j   \left[ \Sigma^i_{\bf v,\hat{m}} ,  {\bf \hat{n}} \cdot {\bm\sigma} \right]_{-} 
  \right) - 
\left( \epsilon^{jkl} {\hat n}^i q^k + \epsilon^{ikl} {\hat n}^j q^k \right) {\bf \hat{n}} \cdot {\bm\sigma}  \left[ \Sigma^l_{\bf v,\hat{m}} ,  {\bf \hat{n}} \cdot {\bm\sigma} \right]_{-} 
\biggr]   \nn
&\times&
J \left[ ({\bf \hat n \cdot q})^2 - {\bf q}^2\right] 
 \biggr\}  \, e^{ - i {\bf q}\cdot {\bf  r}}
  \nn
 \!\! \! &+ &  \!\!  \frac{R}{2}\!\!\int\!\!\! \frac{d{\bf\hat{ m}} \, d {\tilde P'} \tilde {P}'^2  d^3 {q}}{(2\pi)^3}  \big|\tilde\phi \big(
{\bf\tilde P} ' \big) \big|^2 \,
\frac{q^j {q}^i + \delta^{ij} \left(   ({\bf \hat n\cdot q})^2\!  -\! {\bf q}^2  \right)}{2   \gamma ( \tilde {P}' + { v}  \tilde {P}_\parallel ' ) }   \left(\!1 \! 
- \!\frac{i }{2 m }  \epsilon^{l n k} \sigma^k q^l \, {\hat n}^n \! \right)\!  D\!\left[ ({\bf \hat n \cdot q})^2\! -\! {\bf q}^2  \right]\!  e^{ - i {\bf q}\cdot {\bf  r}}\!+\!\text{Rest}.
\label{rhoint3boosted12}
\eea
The expressions in Eq.~(\ref{rhoint3boosted12}) simplify considerably in two extreme
cases. First, in the particle's ZAMF, with $v = 0$, $\gamma =
1$ and  $ D \left[ W\left( \Lambda_{\bf 0}, {\bf\hat { m}}
     \right) \right] = {\rm identity}$, we have $  {\bf n} \big({\bf v},  {\bf\hat { m}} \big) = {\bf\hat 
  m}$, so that we can replace the
integration measure $d {\bf\hat m}$ by  $d {\bf\hat n}$.  

Next, in the infinite-momentum frame (IMF) with $v \to 1$ and $\gamma
\to \infty$, the vector-valued function ${\bf\hat n} $ turns to $\bf{\hat
v}$, so that the integrands in Eq.~(\ref{rhoint3boosted12})  depend on 
${\bf\hat  m}$ only through the Wigner rotation matrices. 
We use the spherical symmetry of $\tilde\phi \big(
{\bf\tilde P} ' \big)$ and factor out the integration over $\tilde P'$  in Eq.~(\ref{rhoint3boosted12}).
To carry out the remaining angular integration over ${\bf\hat m}$ in
spherical coordinates we align the $z$- and $x$-axes along the ${\bf v}$
and ${\bf q}_\perp$ directions, respectively, and denote $\eta =
\cos \theta $. Finally we obtain the following expressions for spatial densities in the IMF:
\be
\label{RhoBoostedCoord12} 
t^{\mu\nu}_{\bf v} ({\bf r})= N_{\phi,R} \int \frac{d^3 {q}}{(2\pi)^3}\, \bar t^{\mu\nu} \left(
  q_\parallel,   {q}_\perp \right)  \, e^{-i {\bf q}\cdot {\bf  r}} + N_{\phi,R,2} \int \frac{d^3 {q}}{(2\pi)^3}\, \bar t_2^{\mu\nu} \left(
  q_\parallel,   {q}_\perp \right)  \, e^{-i {\bf q}\cdot {\bf  r}} +\text{Rest}\,,
\ee
with
 \begin{align}
\bar t^{\mu\nu} \left( q_\parallel,   {q}_\perp \right)  & =   4\pi \gamma \, 
 \left\{ \hat v^{\mu} \hat v^{\nu}  
\left( 1 + \frac{i}{2 m}\, \epsilon^{ilk} \hat v^{i} \sigma^k q^l  \right) 
                             A\left[ - {\bf q}_\perp^2 \right]   
+ \frac{i}{4 m} \biggl( 
\hat v^{\mu}({\boldsymbol \sigma}_\perp\times{\bf q})^\nu+\hat
                                                                                        v^{\nu}({\boldsymbol \sigma}_\perp\times{\bf q})^\mu\right. \nonumber\\ & \left. {} +
(\delta^{\mu 0}\hat v^{\nu}+\delta^{\nu 0}\hat v^{\mu}){\bf \hat v}\cdot({\boldsymbol \sigma}_\perp\times{\bf q})
   \biggr)  
J\left[ - {\bf q}_\perp^2 \right] 
 \right\}   \,,
 \nn  
\bar t_2^{\mu\nu} \left(
  q_\parallel,   {q}_\perp \right)   &  = \frac{\pi}{\gamma}  \,  \alpha  
\left( {q}_v^\mu {q}_v^\nu + g^{\mu\nu} {\bf q}_\perp^2\right) 
 \left( 1 + \frac{i }{2 m } \, \epsilon^{i l k} \hat v^{i} \sigma^k q^l \right) 
D\left[ - {\bf q}_\perp^2
 \right]  \,,  
\label{rhoint3boostedIMFGR}
\end{align}
where $({\boldsymbol \sigma}_\perp\times{\bf q})^0=0$ and $\alpha$ is defined in Eq.~(\ref{defdiv}). 
 Substituting Eq.~(\ref{rhoint3boostedIMFGR}) in Eq.~(\ref{RhoBoostedCoord12}) we obtain, analogously to Eq.~(\ref{defFFr}), expressions for the 
 two-dimensional spatial distributions in the IMF.

\section{Interpretation of the results} 
 \label{interpretation}
 
 We are now in the position to discuss the physical interpretation of the new expressions obtained in the current work.  
  It is clear from Eq.~(\ref{RhoBoostedCoord1RPPR}) that in the IMF, the matrix elements of the EMT in localized states represent 
 two-dimensional distributions in the plane perpendicular to the velocity of the reference frame. 
 Analyzing Eqs.~(\ref{RhoBoostedCoord1}) and (\ref{RhoBoostedCoord2GR}) we conclude that there are two types of 
 contributions: those depending on the velocity of the reference frame and thus characterizing the movement of the 
  system as a whole and the other contributions which are related to internal properties of the system under consideration. 
The contribution generated by $\bar t^{\mu\nu}$ corresponds to the motion of the system as a whole, while the term generated by 
  $\bar t_2^{\mu\nu}$ 
 is not related to this motion and, therefore, should be interpreted as related to internal characteristics. Notice that essentially the same 
 argument supporting separate interpretation of different contributions to the matrix elements of EMT has been given previously in Refs.~\cite{Freese:2021mzg,Freese:2022fat}. 

 Integrating Eq.~(\ref {RhoBoostedCoord1}) for the IMF over all possible directions of the frame velocity, it is easily seen that  the
 resulting two terms are proportional to the corresponding terms in Eq.~(\ref{rhoint4G}), i.e.~to the terms which involve the same 
 form factors.\footnote{Notice that while the coefficients of proportionality are different for different terms, the behavior for $R\to 0$ remains the same, 
 i.e.~the first term diverges while the second term vanishes.} 
 Thus, the spatial distributions in the ZAMF, given by  Eq.~(\ref{rhoint4G}), can be understood as an integral over all directions of the IMF velocity. 
 As evidenced by Eq.~(\ref{RhoBoostedCoord1RPPR}),  the spatial distributions are given by two dimensional images of the system in the IMF. The same can be concluded for the spin-1/2 system using equations \eqref{rhoint2NRGR} and \eqref{RhoBoostedCoord12}.
 Thus the three-dimensional ``picture'' of the system emerges as a ``composition'' of its all possible two-dimensional cuts. 

 As mentioned above for the IMF,  the term corresponding to  $\bar t_2^{\mu\nu}$ characterizes the internal structure. Therefore also in the ZAMF the term   
 \be
\label{deft2ijr} 
t^{ij}_{2} ({\bf r}) = N_{\phi,R,2} \, \int \frac{d^2 \hat n  \,d^3 {q}}{(2\pi)^3} \,  
 \left( - q^i q^j + {\bf q}_\perp^2  \delta^{ij} \right)\, 
\Theta_1 \left[ -{\bf q}_\perp^2 \right]    \, e^{-i {\bf q}\cdot {\bf  r}}
\ee
characterizes the distribution of internal forces. Notice that while the normalization of this quantity depends on the wave packet profile function 
its spatial distribution is uniquely determined by the 
EMT form factor.  We identify the traceless and the trace parts via 
 \begin{equation}
 t^{ij}_{2} ({\bf r}) = \left(\frac{r^i r^j}{r^2} - \frac{1}{3}\, \delta^{ij}\right) s(r)+\delta^{ij}p(r)\,,
 \label{t2r}
 \end{equation}   
 where
  \bea
\label{defPands} 
s ({\bf r}) & = &  N_{\phi,R,2}  \int \frac{d^2 \hat n d^3 {q}}{(2\pi)^3}
\left(\frac{{\bf r }^2 {\bf q}^2 -3({\bf r}\cdot {\bf q})^2}{2\, {\bf r}^2}\right) \Theta_1 \left[ -{\bf q}_\perp^2 \right] e^{-i {\bf q}\cdot {\bf  r}} \nonumber\\ &=&  
N_{\phi,R,2} \, \int d^2 \hat n \, r \frac{d}{d{r} } \frac{1}{r} \frac{d}{d{r} }  \left( \delta(r_\parallel) \tilde\Theta_1 \left[ {\bf r}_\perp \right]\right),
\nonumber\\
p ({\bf r}) & = & N_{\phi,R,2}  \int \frac{d^2 \hat n  \,d^3 {q}}{(2\pi)^3} 
 \left(  {\bf q}_\perp^2 -\frac{{\bf q}^2}{3}\right)
\Theta_1 \left[ -{\bf q}_\perp^2 \right]    \, e^{-i {\bf q}\cdot {\bf  r}} \nonumber\\ &=& 
N_{\phi,R,2} 
\int d^2 \hat n \left( \frac{1}{3} \frac{1}{r^2} \frac{d}{d r} r^2 \frac{d}{d r} - \frac{1}{r_\perp^2} \frac{d}{d r_\perp} r_\perp^2 \frac{d}{d r_\perp} \right) \left( \delta(r_\parallel)\tilde\Theta_1 \left[ {\bf r}_\perp \right] \right) 
\,.
\eea
The quantities $s(r)$ and $p(r)$ have been interpreted in Refs.~\cite{Polyakov:2002yz,Polyakov:2018zvc} as the shear force and the pressure, respectively.\footnote{Notice that this interpretation has been questioned recently in Ref.~\cite{Ji:2021mfb}.}
 It is evident from Eq.~(\ref{defPands}) that the overall normalizations of these quantities depend on the packet while their functional 
 form is uniquely determined by the form factor $\Theta_1$. It is not surprising that the normalization factors of the pressure and shear force distributions vanish in the limit of sharply localized states
   as  these functions are related to the variation of the action with respect to the spatial metric $g_{ik}({\bf r})$. This variation corresponds to a change of the location of the system in three-dimensional space,
   which vanishes for  sharply localized states. Notice that the shape of these distributions does
   not depend on the localization of the system and is uniquely determined by the corresponding form factor.

 \medskip
 
 In exact analogy to the case of the spin-0 system, by integrating the IMF density in Eq.~(\ref{RhoBoostedCoord12}) over all possible directions of the frame velocity, we obtain  
 terms involving different form factors of a spin-1/2 system, which are proportional to the corresponding terms in Eq.~(\ref{rhoint2NRGR}).  Thus, analogously to Eq.~(\ref{t2r}), the term 
  \begin{align}
t^{ij}_{2}(s',s, {\bf r})  
 =  \frac{1}{2} \, N_{\phi,R,2} \, \delta_{s's}\,  \int \frac{d^2 \hat{n} \, d^3 {q}}{(2\pi)^3 } \, 
\, \left( q^i q^j - \delta^{ij} {\bf q}_\perp^2 \right) \, D\left(  - {\bf q}_\perp^2 \right)   
 \, e^{-i {\bf q}\cdot {\bf  r}} 
\label{t212}
\end{align}
can be interpreted as the distribution of the shear force and the internal pressure.  

\medskip
   \begin{figure}[tb]
\center{\includegraphics[width=0.45\linewidth]{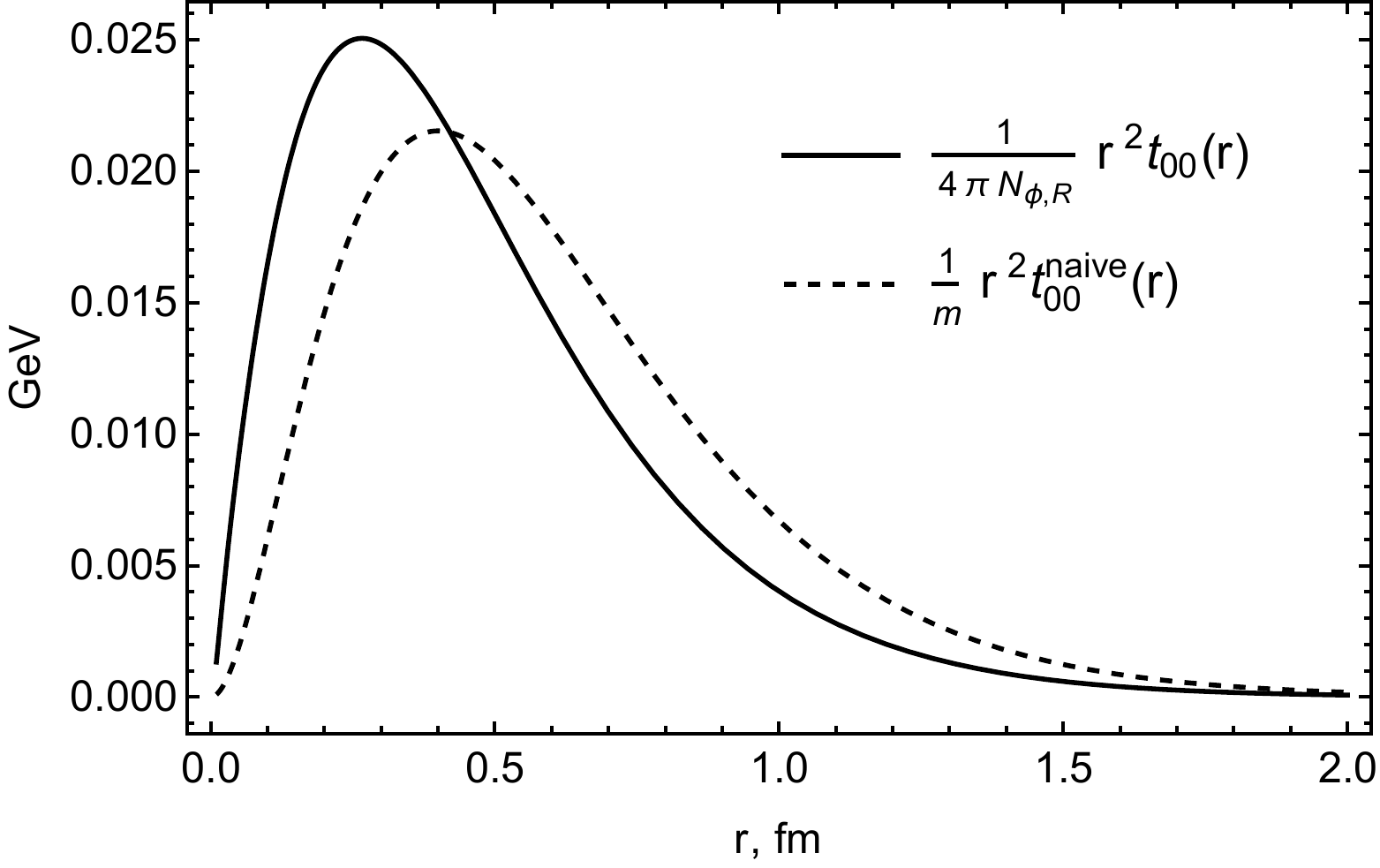}} \\
\vfill
\begin{minipage}[h]{0.495\linewidth}
\center{\includegraphics[width=0.85\linewidth]{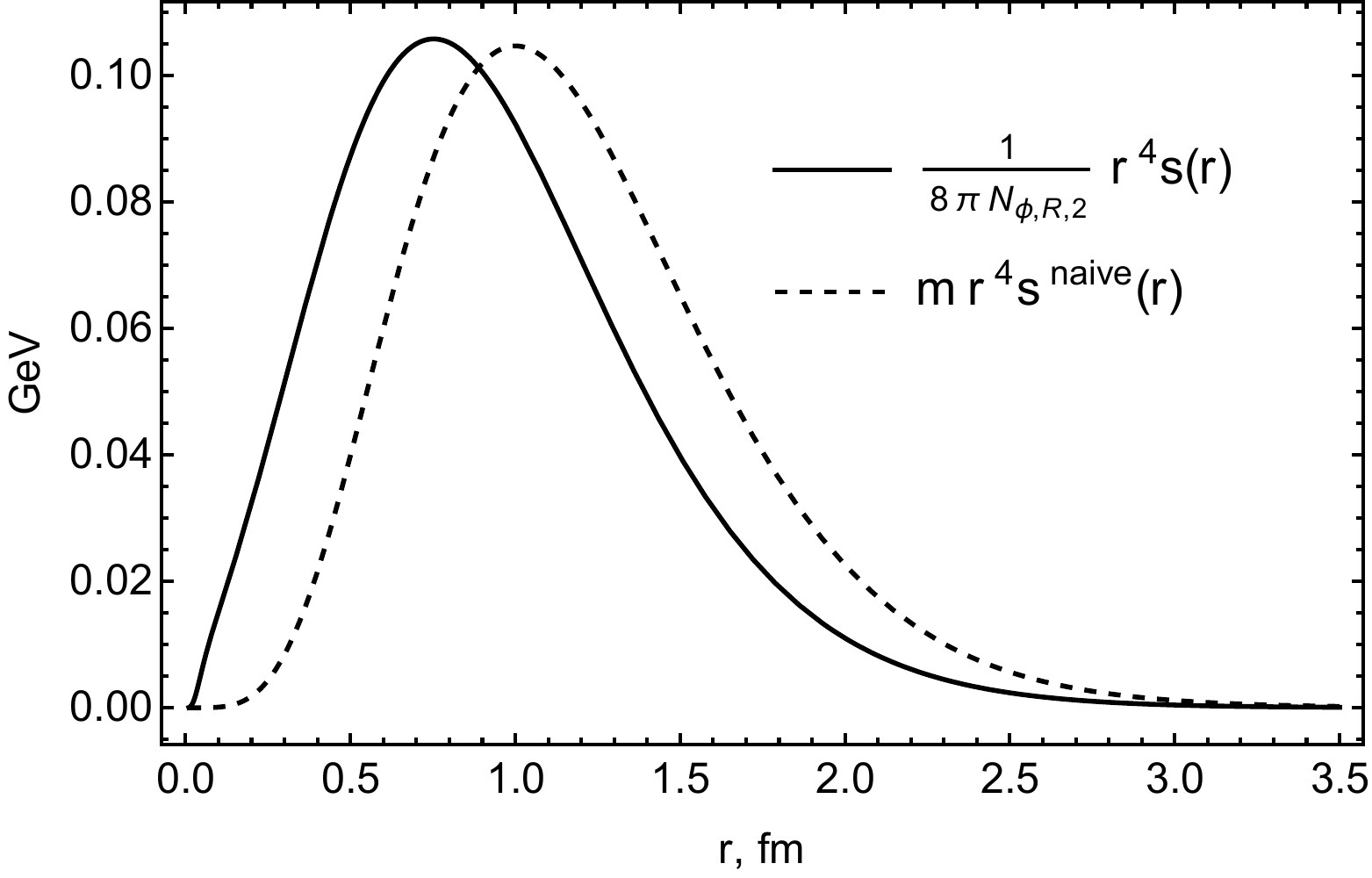}} \\
\end{minipage}
\hfill
\begin{minipage}[h]{0.495\linewidth}
\center{\includegraphics[width=0.85\linewidth]{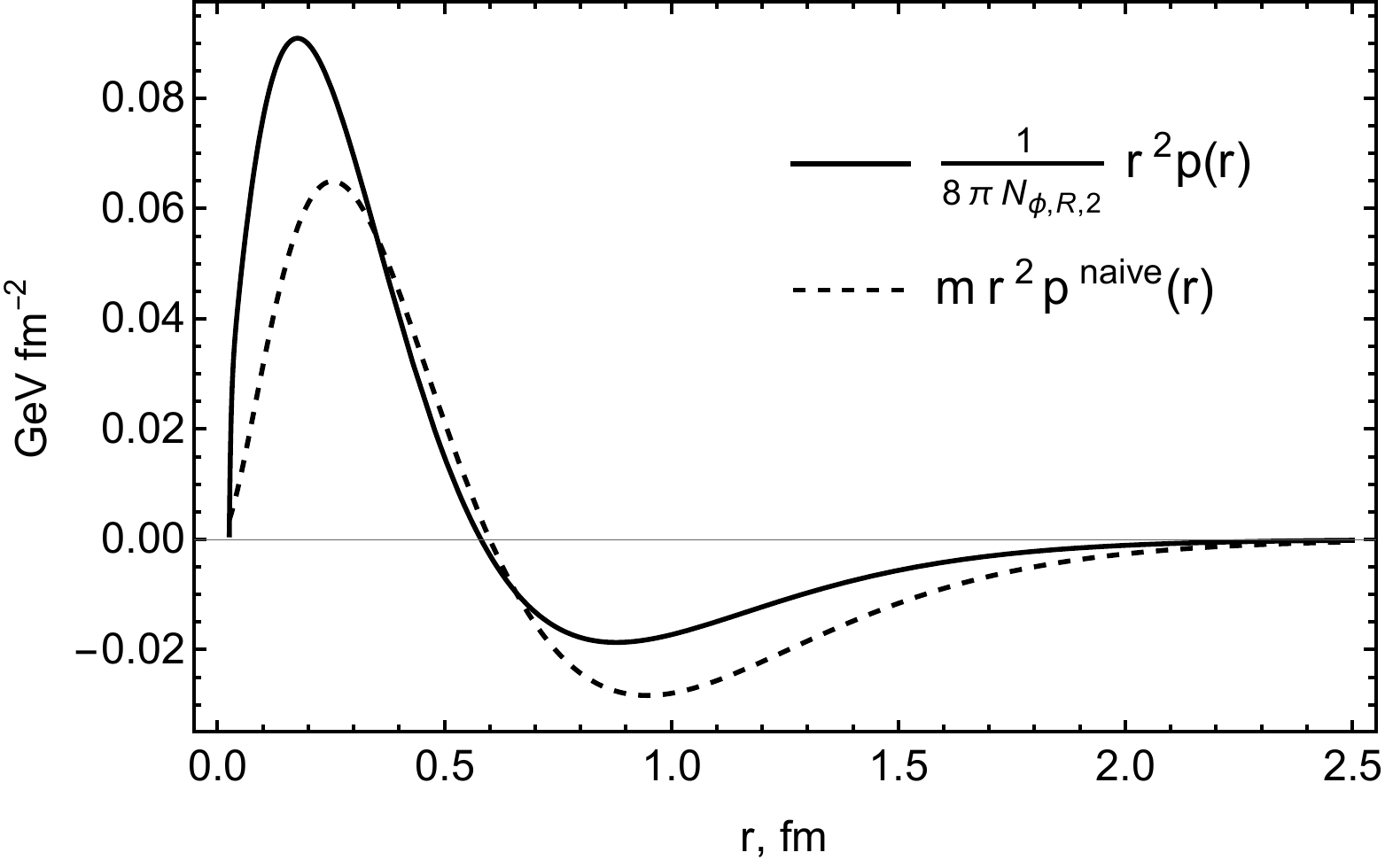}} \\
\end{minipage}
\caption{Normalized energy, pressure and shear force density distributions in the static approximation (dashed lines) and in the ZAMF (solid lines). Normalizations of densities are chosen by dividing the  corresponding matrix elements of the EMT by their integrals over the three dimensional space.   
\label{fig1}
}
\end{figure} 

As mentioned above, a superposition of eigenstates of the energy-momentum vector with different eigenvalues is not an eigenstate of this operator. 
Therefore, in general, the interpretation of the $t^{00} ({\bf r}) $ and $t^{0i} ({\bf r}) $ 
in terms of the energy and momentum distributions seems to be problematic. However, in sharply localized states, where the wave packets are dominated by large momenta, these quantities can be interpreted as energy and momentum spatial distributions, respectively.\footnote{ Notice here that for these terms we do not offer separate interpretation of the $\bar t$ and $t_2$ contributions and therefore the second are dropped as compared to the first.} For a spin-0 system in the ZAMF we have:
 \be
\label{s0deft00} 
t^{00}({\bf r}) =N_{\phi,R} \, \int \frac{d^2 \hat n  \,d^3 {q}}{(2\pi)^3} \,
\Theta_2 \left[ -{\bf q}_\perp^2 \right]    \, e^{-i {\bf q}\cdot {\bf  r}},
\ee
and  for the spin-1/2 system in the ZAMF we obtain the following expressions
  \begin{align}
\label{s1/2deft000i} 
t^{00}({\bf r})& = N_{\phi,R} \, \int \frac{d^2 \hat n  \,d^3 {q}}{(2\pi)^3} \,A \left[ -{\bf q}_\perp^2 \right]    \, e^{-i {\bf q}\cdot {\bf  r}}, \\
t^{0i}({\bf r})& = N_{\phi,R} \dfrac{i\varepsilon^{jkl}}{2m}\, \int \frac{d^2 \hat n  \,d^3 {q}}{(2\pi)^3} \,  \sigma_\perp^k q^l\left(\delta^{ij}+\hat{n}^i\hat{n}^j\right) \,J \left[ -{\bf q}_\perp^2 \right]    \, e^{-i {\bf q}\cdot {\bf  r}}.
  \end{align}
On the other hand, in the static approximation  the states  are defined by packets with  sizes much larger than the 
Compton wavelength of the system.  
 For this case, our expressions of the spatial distributions in the static approximation for a spin-0 system, Eq.~(\ref{rhoint3NRG}), and for a spin-1/2 system,  
Eq.~(\ref{rhoint3RGR}), coincide with the corresponding Breit-frame  
expressions for the $t^{00}$ and $t^{0i}$ components, while for the $t^{ij}$ components the Breit-frame expression reproduces the part describing the internal structure only. 
Thus, the Breit-frame expressions indeed
correspond to the system in the ZAMF in a state described by a packet with the characteristic scale much larger than the Compton wavelength of the system. 
For such a packet the integral is governed 
by momenta much smaller than the mass of the system and, therefore, replacing the 
corresponding energies by the first term in the expansion $E=\sqrt{m^2+{\bf p}^2}=m+{\bf p}^2/(2 m)+\cdots $ is a good approximation. 
That is, the packet is dominated by eigenstates of the energy with the same eigenvalue $m$, and therefore it is also an eigenstate of the energy operator with the eigenvalue $m$. 
Thus,  $t^{00} ({\bf r}) $ can be interpreted in this case as the spatial distribution of the 
mass, which is the same as the full energy of the system in the ZAMF in the static approximation.    
More details of interpretation of the Fourier transforms of the gravitational form factors 
in the Breit frame in terms of  various spatial distributions can be found in Ref.~\cite{Polyakov:2018zvc}. Taking into account  that sharp localization of the system requires a huge amount of energy,
  it is not surprising that the normalization factor for the energy  and the momentum distributions
  $N_{\phi,R}$ explodes. Again, the functional form of these densities is uniquely determined by the
  corresponding form factors. 
  To illustrate the comparison of the static approximation and the ZAMF expressions, in Fig.~\ref{fig1} we plot the normalized energy, pressure and shear force density distributions for a hypothetical spin-0 system with gravitational form factors $\Theta_1(q^2)= 1/(1-q^2/\Lambda^2)^3$ and $\Theta_2(q^2) = 1/(1-q^2/\Lambda^2)^2 $, where we take $\Lambda =1\, {\rm GeV}$.  

Finally, as the spatial densities are given in terms of ${\bf r}= {\bf x} - {\bf X}$, ${\bf X}$ should be interpreted as the position of the center-of-gravity of the system.

\section{Summary and conclusions}
\label{summary}

	To summarize, in the current work we considered the one-particle matrix elements of the EMT in localized states of spin-0 and spin-1/2 systems. 
 Specifying the one-particle states by spherically symmetric wave packets, independent of the spin polarization for the case of spin-1/2 systems,  and sharply localizing system
 we obtained the definition of spatial distributions which are independent of the specific radial form of the packet.    
 This definition is applicable for an arbitrary relative size of the Compton wavelength and other internal characteristics of 
 the system. 
We obtained the expressions for the newly defined quantities in the ZAMF as well as in moving reference frames. 
	Writing the spatial distributions in the ZAMF as integrals of the corresponding expressions in the IMF over all possible directions, we offered a physical interpretation of the obtained results. 

	Finally, using the results of Ref.~\cite{Jaffe:2020ebz}, we also considered the static approximation and obtained the 
spatial distributions, which coincide with the Fourier transforms of the form factors in the Breit frame as originally suggested 
as spatial distributions of various quantities \cite{Polyakov:2018zvc}.

\acknowledgements
We thank H.~Alharazin for helpful discussions.
This work was supported in part by BMBF (Grant No. 05P21PCFP1), by
DFG and NSFC through funds provided to the Sino-German CRC 110
``Symmetries and the Emergence of Structure in QCD'' (NSFC Grant
No. 11621131001, DFG Project-ID 196253076 - TRR 110),
by ERC  NuclearTheory (grant No. 885150), and ERC EXOTIC (grant No. 101018170),
by CAS through a President's International Fellowship Initiative (PIFI)
(Grant No. 2018DM0034), by the VolkswagenStiftung
(Grant No. 93562), by the EU Horizon 2020 research and
innovation programme (STRONG-2020, grant agreement No. 824093),
and by the Heisenberg-Landau Program 2021.


\begin{references}

\bibitem{Hofstadter:1958}
R.~Hofstadter, F.~Bumiller, and M.~R.~Yearian,
Rev. Mod. Phys. {\bf 30}, 482 (1958).

\bibitem{Ernst:1960zza}
F.~J.~Ernst, R.~G.~Sachs and K.~C.~Wali,
Phys. Rev. \textbf{119}, 1105-1114 (1960).

\bibitem{Sachs:1962zzc}
R.~G.~Sachs,
Phys. Rev. \textbf{126}, 2256-2260 (1962).


\bibitem{Polyakov:2002wz}
M.~V.~Polyakov and A.~G.~Shuvaev,
[arXiv:hep-ph/0207153 [hep-ph]].

\bibitem{Polyakov:2002yz} 
  M.~V.~Polyakov,
  Phys.\ Lett.\ B {\bf 555}, 57 (2003). 

\bibitem{Polyakov:2018zvc}
M.~V.~Polyakov and P.~Schweitzer,
Int.\ J.\ Mod.\ Phys.\ A \textbf{33} (2018) no.26, 1830025. 



\bibitem{Burkardt:2000za}
M.~Burkardt,
Phys. Rev. D \textbf{62} (2000), 071503(R),
[erratum: Phys. Rev. D \textbf{66} (2002), 119903(E)].

\bibitem{Miller:2007uy}
G.~A.~Miller,
Phys. Rev. Lett. \textbf{99}, 112001 (2007).

\bibitem{Miller:2009qu}
G.~A.~Miller,
Phys. Rev. C \textbf{79}, 055204 (2009).

\bibitem{Miller:2010nz}
G.~A.~Miller,
Ann. Rev. Nucl. Part. Sci. \textbf{60} (2010), 1-25.


\bibitem{Jaffe:2020ebz}
R.~L.~Jaffe,
Phys. Rev. D \textbf{103} (2021) no.1, 016017.

\bibitem{Miller:2018ybm}
G.~A.~Miller,
Phys. Rev. C \textbf{99}, no.3, 035202 (2019).


\bibitem{Freese:2021czn}
A.~Freese and G.~A.~Miller,
Phys. Rev. D \textbf{103}, 094023 (2021).


\bibitem{Lorce:2020onh}
C.~Lorc\'e,
Phys. Rev. Lett. \textbf{125}, no.23, 232002 (2020),
[arXiv:2007.05318 [hep-ph]].

\bibitem{Lorce:2022cle}
C.~Lorc\'e, P.~Schweitzer and K.~Tezgin,
[arXiv:2202.01192 [hep-ph]].

\bibitem{Guo:2021aik}
Y.~Guo, X.~Ji and K.~Shiells,
Nucl. Phys. B \textbf{969}, 115440 (2021),
[arXiv:2101.05243 [hep-ph]].


\bibitem{Lorce:2018egm}
C.~Lorc\'e, H.~Moutarde and A.~P.~Trawi\'nski,
Eur. Phys. J. C \textbf{79}, no.1, 89 (2019),
[arXiv:1810.09837 [hep-ph]].

\bibitem{Carlson:2022eps}
C.~E.~Carlson,
[arXiv:2208.00826 [hep-ph]].

\bibitem{Freese:2021mzg}
A.~Freese and G.~A.~Miller,
Phys. Rev. D \textbf{105}, no.1, 014003 (2022),
[arXiv:2108.03301 [hep-ph]].

\bibitem{Chen:2022smg}
Y.~Chen and C.~Lorc\'e,
[arXiv:2210.02908 [hep-ph]].

\bibitem{Freese:2022fat}
A.~Freese and G.~A.~Miller,
[arXiv:2210.03807 [hep-ph]].

\bibitem{Panteleeva:2021iip}
J.~Y.~Panteleeva and M.~V.~Polyakov,
Phys. Rev. D \textbf{104} (2021) no.1, 014008,
[arXiv:2102.10902 [hep-ph]].

\bibitem{Epelbaum:2022fjc}
E.~Epelbaum, J.~Gegelia, N.~Lange, U.-G.~Mei\ss{}ner and M.~V.~Polyakov,
Phys. Rev. Lett. \textbf{129}, no.1, 012001 (2022),
[arXiv:2201.02565 [hep-ph]].


\bibitem{Fleming:1974af}
G.~N.~Fleming,  
Physical Reality \& Math. Descrip., 357 (1974). 

\bibitem{Panteleeva:2022khw}
J.~Y.~Panteleeva, E.~Epelbaum, J.~Gegelia and U.-G.~Mei\ss{}ner,
Phys. Rev. D \textbf{106}, no.5, 056019 (2022),
[arXiv:2205.15061 [hep-ph]].


\bibitem{Donoghue:1991qv}
J.~F.~Donoghue and H.~Leutwyler,
Z. Phys. C \textbf{52}, 343-351 (1991).

\bibitem{Kubis:1999db}
B.~Kubis and U.-G.~Mei\ss{}ner,
Nucl. Phys. A \textbf{671}, 332-356 (2000)
[erratum: Nucl. Phys. A \textbf{692}, 647-648 (2001)],
[arXiv:hep-ph/9908261 [hep-ph]].



\bibitem{Gegelia:1994zz}
J.~Gegelia, G.~S.~Japaridze and K.~S.~Turashvili,
Theor. Math. Phys. \textbf{101}, 1313-1319 (1994).

\bibitem{Tanaka:2018wea}
K.~Tanaka,
Phys. Rev. D \textbf{98}, no.3, 034009 (2018),
[arXiv:1806.10591 [hep-ph]].

\bibitem{Tong:2021ctu}
X.~B.~Tong, J.~P.~Ma and F.~Yuan,
Phys. Lett. B \textbf{823}, 136751 (2021),
[arXiv:2101.02395 [hep-ph]].

\bibitem{Tong:2022zax}
X.~B.~Tong, J.~P.~Ma and F.~Yuan,
JHEP \textbf{10}, 046 (2022),
[arXiv:2203.13493 [hep-ph]].

\bibitem{Hoffmann:2018edo}
S.~E.~Hoffmann,
[arXiv:1804.00548 [quant-ph]].


\bibitem{Weinberg:1995mt} 
  S.~Weinberg,
  ``The Quantum theory of fields. Vol. 1, : Foundations,''
  Cambridge University Press (2005-06-02).

%
%
%
%

\bibitem{Ji:2021mfb}
X.~Ji and Y.~Liu,
Phys. Rev. D \textbf{106}, no.3, 034028 (2022),
[arXiv:2110.14781 [hep-ph]].



  \end{references}
\end{document}